\documentclass[11pt]{article}
\usepackage{pst-all}
\usepackage{pstricks}
\usepackage{pstcol,pst-fill,pst-grad}
\usepackage{amsfonts}
\usepackage{graphicx}
\usepackage{amsmath}

\makeatletter \@addtoreset{equation}{section}

\newcommand{\be}{\begin{equation}}
\newcommand{\ee}{\end{equation}}
\newcommand{\bea}{\begin{eqnarray}}
\newcommand{\eea}{\end{eqnarray}}
\begin{document}
\date{}
\title{
\textbf{    Toric Geometry  and String Theory Descriptions  of    Qudit    Systems   }\\
\textbf{   } }
\author{   Adil  Belhaj$^{1}$,     Hamid  Ez-Zahraouy$^{2}$, Moulay  Brahim Sedra$^{3}$
\hspace*{-8pt} \\
\\
{\small $^{1}$D\'epartement de Physique, Facult\'e
Polydisciplinaire, Universit\'e Sultan Moulay Slimane}\\{ \small
B\'eni Mellal, Morocco }
\\  {\small $^{2}$Laboratoire de Magn\'etisme et Physique des
Hautes \'Energies, Facult\'e des Sciences }\\{\small Universit\'e
Mohammed V, Av. Ibn Batouta,  B. P 1014, Rabat, Morocco} \\{\small
$^{3}$LHESIR,   D\'{e}partement de Physique, Facult\'{e} des
Sciences, Universit\'{e} Ibn Tofail }\\{ \small K\'{e}nitra,
Morocco} } \maketitle

\begin{abstract}
In this paper, we  propose a new  way   to approach  qudit systems
using
 toric geometry   and related topics including  the local  mirror symmetry used in  the string theory compactification. We
refer to such systems as  $(n,d)$  quantum systems  where $n$ and
$d$ denote the number of the qudits and the basis  states
respectively.
  Concretely, we first
relate  the  $(n,d)$  quantum systems  to the  holomorphic sections
of line bundles   on $n$ dimensional projective spaces $\bf CP^{n}$
with degree $n(d-1)$. These sections are in one-to-one
correspondence with $d^n$ integral points on  a $n$-dimensional
simplex.   Then, we explore the  local mirror map in the toric
geometry language  to establish a linkage between  the $(n,d)$
quantum systems and type II D-branes placed at singularities of
local Calabi-Yau manifolds. $(1,d)$ and $(2,d)$ are analyzed in some
details and are found to  be related to the mirror of the ALE space
with the $A_{d-1}$ singularity and a generalized conifold
respectively.

\end{abstract}


\thispagestyle{empty}

\newpage \setcounter{page}{1} \newpage

\section{Introduction}
Recently,  many  efforts have been devoted to study   quantum
computation and information theory  using different approaches.
These activities have brought new understanding of the fundamental
physics associated with quantum theories   \cite{1,2,3,4,40,41}.  It
is shown that the quantum bit or the qubit, which is based on   a
superposition of the two possible states,  is considered as   the
most basic building block of such quantum theories.    In fact,
qubit and its generalizations are connected to many theories
including string theory, D-branes, black holes, toric geometry and
supermanifolds \cite{5,6,7,8,9}. A particular emphasis put on the
link with the  black holes and  branes  obtained from supergravity
models in high dimensions. More precisely, a nice interplay between
the STU black holes having eight charges and three qubits have been
proposed and developed  in \cite{6,7}.

More recently, many works  have suggested  extended models using
different ways. In connection  with supersymmetry,  superqubits  get
developed using  $so(2|1)$ Lie superalgebra \cite{10,11}.  On the
other hand, the qudit,   which is   a $d$-level quantum system,
emerge naturally in the extended models by considering more than two
bosonic  states  on which  the qubit is built. It is recalled that
$n$ qudits  can  be characterized by a  couple $(n,d)$ where $n$ and
$d$ indicate the number of the qudits and  the states  respectively.
Here,   we refer to these systems as $(n,d)$ quantum systems.
Inspection shows that  there have been some  attempts to dealt  with
such quantum systems but unfortunately with partial results only.

The aim of the paper is to enrich  these activities by proposing a
new way to deal  the qudits  using   toric geometry and its relation
to type II D-branes placed at singularities of Calabi-Yau manifolds
used  in the string theory compactification.
 In particular,  we
relate  the  $(n,d)$ quantum  systems  to the  holomorphic sections
of  the line bundles  on $n$ dimensional projective spaces $\bf
CP^{n}$ with degree $n(d-1)$. Using toric geometry technics,  these
sections are in one-to-one correspondence with $d^n$ integral points
on $n$-dimensional simplex.   To give an explicit analysis, a
particular interest has been on lower dimensional examples. Then, we
explore the  local mirror map  to elaborate a linkage between the
$(n,d)$ quantum  system and type II D-branes placed at toric
singularities. $(1,d)$ and $(2,d)$ are analyzed in some details and
are found to be related to the mirror of the ALE space with the
$A_{d-1}$ singularity and a  generalized conifold respectively.

The present  paper is organized as follows. In section 2,  we give a
short review on the qubit and qudit systems. Section 3 concerns a
toric description of the $(1,d)$  quantum systems while the
generalization to $(n,d)$ ones is discussed in section 4.  In
section 5, we propose a string theory interpretation in terms of
type II  D-branes placed at toric Calabi-Yau  singularities using
local mirror symmetry. Section 6 contains some including remarks.

\section{Qudit quantum   systems}
Inspired  by toric  varieties  and motivated by the existence of
combinatorial calculations in quantum information, we use toric
geometry  to handle multiple of qudit systems. Concretely, we
elaborate a toric  description  in terms of holomorphic section of
line bundles on the  projective spaces $\bf CP^{n}$. Then, we
propose a stringy interpretation using type II D-branes placed at
mirror Calabi-Yau singularities used in the string theory
compactification. To end, let us first recall the qubit which has
been extensively studied using
  different physical and  mathematical approaches \cite{1,2,3,4}.
\subsection{Qubit quantum   systems}
The qubit  is a two state  system which can be  realized, for
instance,   by   the position of the  electron  in the hydrogen
atom.
 The superposition state of  a single  qubit is generally  given by
 the following
 Dirac notation
\begin{equation}
|\psi\rangle=a_0|0\rangle+a_1 |1\rangle
\end{equation}
 where $a_i$  are complex  numbers   satisfying the normalization
condition
\begin{equation}
|a_0|^2+|a_1 |^2=1.
\end{equation}
It is observed that  this equation can be  interpreted geometrically
in terms of the so called Bloch sphere \cite{1,2,3,4}. The  analysis
can be extended to more than one qubit which has been used to
discuss entanglement states. In fact, the two qubits are four  level
systems. Using the usual
 notation  $|ij\rangle=|i\rangle|j\rangle$,   the corresponding
 state superposition  can be expressed as
\begin{equation}
|\psi\rangle=a_{00}|00\rangle+a_{10}
|10\rangle+a_{01}|01\rangle+a_{11} |11\rangle,
\end{equation}
where $a_{ij}$  are  complex numbers verifying  now the following
normalization condition
\begin{equation}
|a_{00}|^2+|a_{10}|^2+|a_{01}|^2+|a_{11}|^2=1.
\end{equation}
 $n$ qubits, in fact,  are   $2^n$ configuration states   which can be represented by
   graphs sharing a
strong resemblance with
 a particular class of  Adinkras  formed
by $2^n$ nodes connected  with
 $n$ colored lines \cite{9}.  This observation has been explored  to show a new  similarity between
  $n$ qubit systems and $2^n$ cycles embedded  in $n$-dimensional torii $T^n$. In this way, a
quantum state has been   interpreted   as the Poincar\'e dual of the
real homology cycle in  $T^n$  on which  type II D-branes can wrap
to generate black holes  in  the compactification of type IIA
superstring on $T^n$.

\subsection{Qudit  systems}
As in  the case of qubits,  the qudits represent, for instance, a
physical system of  spin $\frac{d-1}{2}$ particles  with $d$ states.
In the basis formed by the  states
$\{|i\rangle,\;i=0,1,\ldots,d-1\}$, the general state can be written
as follows
\begin{equation}
|\psi\rangle=a_0|0\rangle+a_1 |1\rangle+\ldots+ a_{d-1} |d-1\rangle,
\end{equation}
 where $a_i$  are complex  numbers  satisfying now  the extended   normalization
condition
\begin{equation}
\sum \limits_{i=0}^{d-1}|a_i|^2=1.
\end{equation}
For  the organization reason, one may be interested in
characterizing the multiple of qudits. In fact,   it  should be
evident  to characterize them by a couple $(n,d)$ where $n$ and $d$
indicate the number of the qudits and  the states respectively.
Using  a similar qubit notation, the general state can take  the
form
\begin{equation}
\label{qudit} |\psi\rangle=\sum\limits_{i_1\ldots
i_n=0,1,\ldots,d-1}a_{ i_1\ldots i_n}|i_1 \ldots i_n\rangle.
\end{equation}

In the handling of such physical quantities, it  is remarked that
there are some similarities   between qudits and toric geometry.
Here, thought, we will be concerned with them.  Our main objective
is to explore the toric geometry  language,  and related topics
including string theory compactification,  to deal with the $(n,d)$
quantum systems.

\section{Toric geometry realization  of  a single qudit}
In this section, we borrow the idea of toric geometry to discuss
qudits and make contact with other physics including the
compactification of higher dimensional theories.  We thus  are
considering a basic correspondence by replacing states by
holomorphic sections of line bundles on toric varieties.  Before
going ahead,   we first start by giving some basic facts on  toric
geometry being  one of the most useful mathematical tools used in
 superstrings, M and F-theories \cite{12,13,14,15}. Indeed,
 a $n$-dimensional toric variety $V^n$  is a complex manifold which
   can be
represented by a toric graph  known by  polytope $\Delta(V^n)$
consisting of $n+r$ vertices $v_i$ in an $Z^n$ lattice satisfying
\begin{equation}
\label{toric} \sum\limits_{i=0}^{n+r-1} q_i^av_i=0, \qquad
a=1,\ldots,r
\end{equation}
where $q_i^a$  are  called Mori vectors.  It is observed that $V^n$
can be  fixed by $\{ q^a_i, v_i\}$  toric data. It is interesting to
note, in passing,  that the local Calabi-Yau  condition is satisfied
by the following relation
\begin{equation}
 \sum\limits_{i=0}^{n+r-1} q_i^a=0.
\end{equation}
Familiar  examples of toric varieties, which have  many applications
in both mathematics and quantum physics,  are  the   complex
projective spaces $\bf CP^n$\cite{16,17}. For $\bf CP^n$, $q_i^a$
reduce  to a simple vector $q_i=(1,\ldots,1)$ with $a=1$ and the
equation (\ref{toric}) becomes a simple relation given by
\begin{equation}
v_0+\ldots+v_{n}=0
\end{equation}
forming a $n$-dimensional simplex which will be  explored    to
discuss holomorphic sections of  the line bundles on $\bf CP^n$.

Having introduced the mathematical backgrounds, we move now to
present a  toric description of  a single qudit.  For a sake of
simplicity, we consider quantum trit or qutrit associated with the
couple $(n,d)=(1,3)$. It is a three level system which can be
simulated by
 a physical  system dealing with  spin-1 particles.  It can be also realized by
 the  degree of freedom of a photon living in
five  dimensional space-time, which can be obtained from the string
theory compactification on a 5-dimensional compact sapce. In this
way, a superposition state is a vector of a three dimensional
Hilbert space, given by
\begin{equation}
\label{state1} |\psi\rangle=a_0|0\rangle+a_1 |1\rangle+a_2
|2\rangle,
\end{equation}
where $|i\rangle,\; (i=0,1,2)$,   denote its  basis.  $a_i$  are
complex coefficients  satisfying the normalization condition
\begin{equation}
\label{cp2} |a_0|^2+|a_1 |^2+|a_2|^2=1,
\end{equation}
where $|a_i |^2$ is the probability of measuring the qutrit in the
state $|i\rangle$.  Equation (\ref{cp2})  plays the  same role as
the Bloch sphere in the single qubit model associated with the
$(n,d)=(1,2)$ quantum  system. To give a toric description of the
qudit, the state $|i\rangle$ is replaced by a    Laurent monomial
$z^i$ as follows
\begin{equation}
|i\rangle \to z^i, \qquad i=1,2,3.
\end{equation}
We naturally requires that (\ref{state1})  takes the following form
\begin{equation}
\label{pol13} |\psi\rangle \to P(z)=a_0+a_1z+a_2z^2,
\end{equation}
where $z$ will be  interpreted  as the complex  coordinate of one
dimensional projective space ${\bf CP}^1$.  It recalled that ${\bf
CP}^1$  is the simplest example in toric geometry which turns out to
play a primordial  role in  the elaboration of our basic
correspondence. It is considered a building block of higher
dimensional toric manifolds explored in string theory
compactification in particular in the blow up of  ADE singularities
of  Calabi-Yau manifolds producing four dimensional quantum field
models \cite{12,13,14,15}. It is known that  ${\bf CP}^1$ has a U(1)
toric action having two fixed points $p_1$ and $p_2$ describing
respectively north and south poles of the real two sphere,
identified with $\bf CP^1$. The corresponding toric graph is one
dimensional polytope which is just the 1-dimensional simplex
identified with
 the interval $[p_1, p_2]$.  Using  Laurent polynomials, (\ref{pol13}) can be associated with holomorphic
sections of  line bundles on ${\bf CP}^1$.  It is observed  that
such line  bundles   are associated with integral points on the
interval \cite{17}. In this way,   the number of holomorphic
sections of the bundle  corresponds  to the number of integral
points on the interval. Indeed, let us assume that the interval
$[p_1, p_2]$ goes from 0 to 2. The  holomorphic sections can be
identified now  with the terms $z^i$, where $i=0,1,2$.   Each
integral point, specified by  $i$ on the interval $[0, 2]$,
corresponds to a section of the line  bundle. The  state
$|i\rangle$, appearing in the $(1,3)$ quantum system,   is
associated  then with   such a holomorphic section.

It should be evident that this extends to  the    $(1,d)$ quantum
system. In this case, the general state can be represented by the
following polynomial
\begin{equation}
|\psi\rangle \to P(z)=a_0+a_1z+\ldots+ a_{d-1}z^{d-1}.
\end{equation}
 This should be associated with a 1-dimensional simplex   involving  $d$  integral points. With this analogy,
we are considering a similarity between holomorphic sections of line
bundles on  $\bf CP^1$ with degree $d-1$ and  $(1,d)$ quantum
system. This  correspondence can  be illustrated in figure 1.
\begin{center}
\begin{figure}[!ht]
\begin{center}
 {\includegraphics[width=10cm]{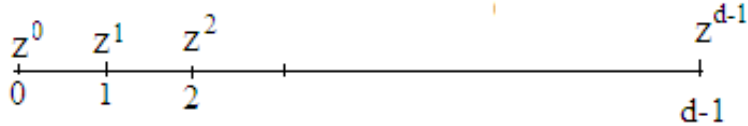}}
 \caption{ Toric realization of  the $(1, d)$ quantum system. } \label{fig1:fig2x} \vspace*{-.2cm}
   \end{center}
\end{figure}
\end{center}
\section{Multiple qudit systems}

We  can also discuss, from the perspective of  toric geometry, the
more general case  associated with the  $(n,d)$ quantum system. The
analysis  requires the  incorporation of  the $n$ dimensional
projective space ${\bf CP}^n$ and  holomorphic sections of line
bundles on it. These sections will  be  controlled by $d^n$ integral
points placed  at  a $n$-dimensional simplex representing the  toric
description of the line bundles on ${\bf CP}^n$ with degree
$n(d-1)$.

The general graphic representation is beyond the scope of the
present work, though we will consider  an  explicit example
corresponding to  $(n,d)=(2,d)$. Then, we give some remarks on  the
general case.  For $(n,d)=(2,d)$, (\ref{qudit}) reduces to
\begin{equation}
|\psi\rangle=\sum\limits_{i_1i_2=0,1,\ldots,d-1}a_{
i_1i_2}|i_1i_2\rangle.
\end{equation}
Using  the above basic   correspondence, this state can be
represented by the following polynomial with two complex variables
$z_1$ and $z_2$
\begin{equation}
|\psi\rangle \to P(z_1,z_2)=\sum\limits_{i_1i_2=0,1,\ldots,d-1}a_{
i_1i_2}z_1^{i_1}z_2^{i_2}.
\end{equation}
Here  $z_1$ and $z_2$  can be  considered as   local  complex
variables defining ${\bf CP}^2$. It is worth noting that the number
of the complex variables is exactly the number of qudits ( here
$n=2$). Roughly speaking, ${\bf CP}^2$ is a complex two dimensional
toric manifold with an U(1)$^2$ toric action exhibiting three fixed
points $p_1$, $p_2$ and $p_3$. The corresponding polytope  is  a
2-dimensional simplex belonging to the $Z^2$ square lattice.  It is
obtained from the intersection of three ${\bf CP}^1$ complex lines
defining  a triangle ($p_1,p_2,p_3$) in toric geometry language. The
holomorphic sections of the line bundles on ${\bf CP}^2$ are
characterized by integral points on the triangle ($p_1,p_2,p_3$).

To make contact with toric geometry,  we should  regard  $(i_1,i_2)$
as integral points placed on a 2-dimensional simplex.  Then, we
assign  them to a monomial $z_1^{i_1}z_2^{i_2}$. In toric geometry
language, this monomial can be interpreted as a holomorphic section
of the  line bundles on ${\bf CP}^2$ \cite{17}. Let us illustrate
this model for   $(n,d)=(2, 3)$. The corresponding general state of
this qudit
 reads
\begin{equation}
|\psi\rangle=a_{00}|00\rangle+a_{10}
|10\rangle+a_{01}|01\rangle+c_{11} |11\rangle+
a_{20}|20\rangle+a_{02} |02\rangle+a_{21}|21\rangle+a_{12}
|12\rangle+ a_{22} |22\rangle
\end{equation}
which can be replaced by the following Laurent polynomial
\begin{equation}
|\psi\rangle \to P(z_1,z_2)=\sum\limits_{i_1i_2=0,1,2}a_{
i_1i_2}z_1^{i_1}z_2^{i_2}.
\end{equation}
The integers  $i_1$ and $i_2$ should satisfy the following
constrains
\begin{equation}
0\leq i_1\leq 2,\quad 0\leq i_2\leq 2, \quad  0\leq i_1+i_2 \leq 4,
\end{equation}
producing  holomorphic sections on  the line bundles on  ${\bf
CP}^2$ associated with the  $(2, 3)$ quantum system. They represent
a collection of $3^2=9$ integral points in ${\bf Z}^2$ given by
\begin{equation}
\{(0,0),(1,0),(0,1),(1,1),(2,0),(0,2),(1,2),(2,1),(2,2)\}.
\end{equation}
belonging  to  a  2-dimensional simplex representing  a toric
realization of  the line bundles on ${\bf CP}^2$ with degree 4, as
illustrated in figure 2.

\begin{center}
\begin{figure}[!ht]
\begin{center}
 {\includegraphics[width=8cm]{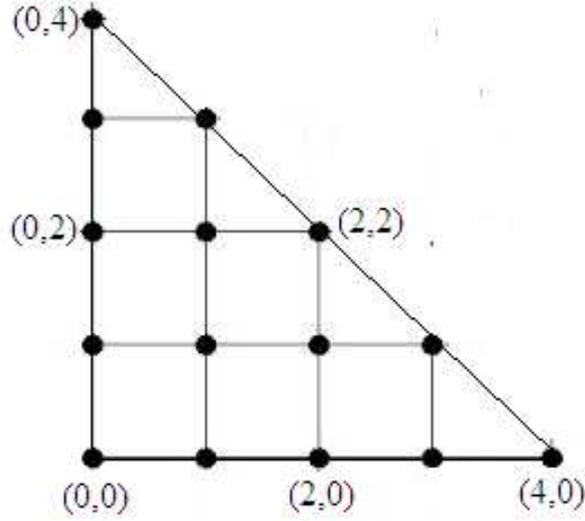}}
 \caption{  Toric realization of  the $(2, 3)$ quantum system. } \label{fig1:fig2x} \vspace*{-.2cm}
   \end{center}
\end{figure}
\end{center}

For  a physical system with $d$ states, the number $i_1$ and $i_2$
have to be lower  than a maximal value in terms of $d$. These
numbers are  seen to satisfy this condition
\begin{equation}
0\leq i_1\leq d-1,\quad  0\leq i_2\leq d-1
\end{equation}
with the extra constraint
\begin{equation}
  0\leq i_1+i_2 \leq 2(d-1).
\end{equation}
The pairs $(i_1,i_2)$ generate now    a  toric description of
holomorphic sections of the  line bundles  ${\bf CP}^2$  with degree
$2(d-1)$ associated with the  $(2,d)$ quantum system. It is
represented by $d^2$ integral points on a  2-dimensional simplex.

These results  can be extended to higher orders of qudits. More
precisely, we can do something similar for the more general case
with  $n > 2$ corresponding to the $(n,d)$  quantum system using $n$
local complex variables defining ${\bf CP}^n$. The main point is to
find all possible integral points of the $n$-dimensional lattice
${\bf Z}^n$ corresponding to  the monomials
\begin{equation}
z_1^{i_1}z_2^{i_2}\ldots z_n^{i_n}.
\end{equation}
Quantum information requires that  $(i_1,\ldots,i_n )$ should
satisfy
\begin{equation}
0\leq i_\ell \leq d-1,\quad i_\ell=1,\ldots, n
\end{equation}
together  with the extra constraint
\begin{equation}
  0\leq  \sum \limits_{\ell=1}^{n} i_\ell \leq n(d-1)
\end{equation}
 producing holomorphic sections on  the  line bundles on  a ${\bf CP}^n$   with degree $n(d-1)$.
 These  $d^n$ integral points give
 a toric description of  the $(n,d)$  quantum system on a
$n$-dimensional simplex.

\section{String theory  interpretation}
Toric geometry has been  considered as a powerful tool to study
mirror symmetry in the context of supersrtring theory
compactification on global and local Calabi-Yau manifolds with
singularities in the presence of  type II  D-branes \cite{18,19}. On
the basis on these activities, we bridge  qudits  to  brane
configurations which are mirror to the singularities of local
Calabi-Yau manifolds having toric representations. In fact, it has
been suggested  that toric geometry  can encode information
concerning  D-brane configurations in type II superstrings
\cite{17}. This can be done by exploring the  toric graph to encode
the corresponding physical data. We will see that this may produce a
new link between $n$ qudits and D-branes moving on local Calabi-Yau
geometries. Given a toric a toric manifold, the local mirror
symmetry maps the toric data $\{q^a_i,v_i \}$ to an algebraic
equation describing the mirror geometry given by
\begin{equation}
\label{mirror} \sum_{i=0}^{n-1}a_iy_i=0,
\end{equation}
where $a_i$  are  complex coefficients  and  $y_i$  monomials
satisfying the famous  mirror constraint relations
\begin{equation}
\prod_{i=0}^{n-1}y_i^{q^a_i}=1, \qquad a=1,\ldots,r
\end{equation}
There are many ways to   solve these constraint  equations giving
the  type II local mirror geometries \cite{18,19}. These geometries
have been extensively studied for bosonic and fermionic Calabi-Yau
manifolds in connection with sigma model with four supercharges
\cite{20,21}.

In what follows,  we will show that  the  $(n,d)$  quantum system
can be related with  the local  mirror geometry equations. For the
simplicity, we consider lower dimensional models  but  we expect
that possible generalizations  to  higher dimensional cases could be
done using a similar method. Indeed, we consider  the case of
$(n,d)=(1,d)$. In fact, this model can be related to the mirror
symmetry of M-theory on ALE space with $A_{d-1}$ singularity given
by the following algebraic equation
\begin{equation}
\label{A1} x_1x_2=x_3^d
\end{equation}
where $x_1$,  $x_2$ and $x_3$  are complex variables \cite{12}. This
equation involves  a  singularity  located at the $C^3$ origin
$x_1=x_2=x_3=0$. The latter  can  be  resolved in two ways either by
deforming the complex structure of (\ref{A1})  or  varying its
Kahler structure. These  two deformations are equivalent due to the
self mirror property of the ALE spaces considered as local versions
of the K3 surfaces\cite{12}. Using sigma model approach,  it is
shown that the Kahler deformation consists on replacing the singular
point by a collection of ${\bf CP}^1$ according to the Dynkin
diagram of the  $A_{d-1}$  finite Lie algebra.  This nice connection
between toric geometry  and Lie algebras  will be explored to build
a bridge between local mirror geometries and  the $(n,d)$ quantum
systems.

 Roughly speaking,  the above
model is dual to the mirror of  type IIA superstring with $d$
D6-branes.  Identifying $q^a_i$, up some details,  with the Cartan
matrix of  the finite Lie algebra $A_{d-1}$,  the mirror geometry
equations of  the  ALE space  with $A_{d-1}$ singularity  read as
\begin{equation}
y_iy_{i+2}=y^2_{i+1}, \qquad i=0,\ldots, d-1.
\end{equation}
These equations can be solved  by the following monomials
\begin{equation}
 y_i=z^{i}
\end{equation}
 which can be identified with  the holomorphic sections of the line  bundles on   ${\bf
CP}^1$ as we have discussed previously.  Using  mirror equations
(\ref{mirror}), the polynomial representation of the  $(1,d)$
quantum information system can be identified  with the mirror
geometry solution given by
\begin{equation}
\sum\limits_{i=0}^{d-1}a_{ i}z^{i}=uv
\end{equation}
where $u$ and $v$ are auxiliary variables which   have been
introduced to recover the right dimension. It is worth noting that
the quadratic term  $uv$ has no physical importance since it does
not affect the modouli space of the mirror geometry. The relevant
physical  quantities are  $a_i$  specifying the  D6-brane locations
  associated with   $d$ integral   points  on a  1-dimensional simplex as illustrated in figure
  1.  In this scenario, we
associate  the holomorphic sections of the line   bundles on  ${\bf
CP}^1$ with D6-branes in type IIA superstring.

The second example we want to discuss is the  $(2,d)$ quantum
system. In this case, the  $(i_1,i_2)$  can be associated with the
following  mirror geometry equation
\begin{equation}
\sum\limits_{i_1i_2=0,1,\ldots,d-1}a_{ i_1i_2}z_1^{i_1}z_2^{i_2}=uv
\end{equation}
where $(i_1,i_2)$ have been explored to solve the  mirror equations.
A close  inspection shows that    this equation describes the local
mirror geometry of a generalized conifold  given by
\begin{equation}
 x_1x_2=x_3^dx_4^d,
\end{equation}
where $x_1$,  $x_2$,  $x_3$ and $x_4$  are complex variables.  This
geometry has been  dealt  with  in connection with D-branes placed
at  toric singularities of local Calani-Yau manifolds  as developed
in \cite{22,23}. Inspired by these activities, the  $(2,d)$ quantum
system  can be related to the D3-branes
 probing  such a geometry. Indeed, the complex  parameters  $a_{ i_1i_2}$ can be interpreted
as  the moduli space of the gauge theory living on the world volume
of  D3-branes. In this way, $(i_1,i_2)$ give the locations of  the
D3-branes as  illustrated in figure 2. According to \cite{22}, this
model can be T-dual with D4-branes  wrapping $S^1$ and stretched
between NS-branes placed at various points on $S^1$. We believe that
this connection deserves more deeper study. We  hope to come back to
this issue in future.
\section{Conclusion}
In this paper, we  have   approached   qudit systems using
 toric geometry language  and its relation  with type II D-branes placed at
Calabi-Yau singularities explored   in string theory
compactification. We have refereed to  these  systems   as $(n,d)$
quantum   systems where $n$ and $d$ indicate the number of the
qudits and the states respectively.
  Concretely, we have shown that
   the  $(n,d)$ quantum  systems  can be related  to the  holomorphic sections of   the line
bundles on $n$ dimensional projective spaces $\bf CP^{n}$.  These
sections are in one-to-one correspondence with $d^n$ integral points
on $n$-dimensional simplex associated with the toric realization of
holomorphic section of line bundles with degree $n(d-1)$.  Using the
local mirror map in the toric language, we have established    a
linkage between the  $(n,d)$  quantum systems and type D-branes
placed at singularities of local Calab-Yau manifolds used in the
type II superstring compactification. It has been found that $(1,d)$
and $(2,d)$ can be linked with the mirror of the ALE spaces with
$A_{d-1}$ singularity and  a generalized conifold respectively.

We expect that our  approach can be  adaptable to a broad variety of
geometries represented by non trivial polytopes going beyond the
simplex geometry. We anticipate that other concepts used in quantum
information could be discussed by implementing  extra  constraints
on the integral points  belonging on such  polytopes.

 We believe that the present   study  could  be considered as a
first steep for developing such concepts using toric geometry
technics used in the  string theory compactification.  Concretely,
we intend to discuss elsewhere  gates and non separated states in
future works.

\end{document}